\documentclass[preprints,article,accept,moreauthors,pdftex]{Definitions/mdpi}
\firstpage{1}
\pubvolume{1}
\issuenum{1}
\articlenumber{0}
\pubyear{2022}
\copyrightyear{2022}
\datereceived{}
\dateaccepted{}
\datepublished{}
\hreflink{https://doi.org/} 

\Title{Absorption spectra of the purple nonsulfur bacteria light-harvesting complex: a DFT study of the B800 part}

\TitleCitation{Absorption spectra of the purple nonsulfur bacteria LH2: a DFT study of the B800 part}

\Author{L.V. Begunovich$^{1}$\orcidA{}, E.A. Kovaleva$^{2,*}$\orcidB{}, M.M. Korshunov$^{1,2,}$\orcidC{}, and V.F. Shabanov$^{1,2}$}

\AuthorNames{L.V. Begunovich, E.A. Kovaleva, M.M. Korshunov, and V.F. Shabanov}

\AuthorCitation{Begunovich, L.V.; Kovaleva, E.A.; Korshunov, M.M.; Shabanov, V.F.}

\address{%
$^{1}$ \quad Federal Research Center KSC SB RAS, Akademgorodok, 660036, Krasnoyarsk, Russia\\
$^{2}$ \quad Kirensky Institute of Physics, Federal Research Center KSC SB RAS, Akademgorodok, 660036, Krasnoyarsk, Russia}

\corres{Correspondence: kovaleva.evgeniya1991@mail.ru}

\abstract{We've studied the B800 part of \textit{Rhodoblastus acidophilus} light-harvesting  complex (LH2) by several quantum chemical techniques based on the density functional theory (DFT) and determined the specific method and a minimal reliable model suitable for further studies of the LH2. In addition to bacteriochlorophyll \textit{a} molecules, the minimal model includes two $\alpha$ and one $\beta$ chain amino acids. Within the model, we are able to reproduce the contribution of the B800 ring of nine bacteriochlorophyll \textit{a} molecules to the near infrared $Q_y$ absorption band. We also discuss the use of hybrid DFT calculations for precise energy and optical estimations and DFT-based tight binding (DFTB) method for the large-scale calculations. Crucial importance of Hartree-Fock exchange interaction for the correct description of B800 peak position was shown.}


\begin{document}

\section{\label{Intro} Introduction}

Photosynthesis is the most important life-supporting process on Earth. It consists of several steps starting from the absorption of light and resulting in production of carbohydrates. The steps have different timescales ranging from peto- and femtoseconds to seconds and the whole process involves spatial scale from nanometers (chlorophyll molecule) to meters (canopy)~\cite{Capretti2019,Chow2003}. While the description of the longer steps may belong to a working space of physical chemistry and biochemistry, shorter steps require quantum physical interpretation~\cite{Yakovlev2016}. The very beginning, absorption of light by the so-called light-harvesting complex (LHC) is extremely effective since the millions of years of evolution and natural selection left only the most optimized structures. The quantum efficiency of the charge separation is about 100\%. There are only a few types of molecules at the heart of the LHCs. Those are variations of chlorophyll, carotenoid, and phycobilin pigments~\cite{Mirkovic2017}. They may be combined in different structures resulting in a variety of plants, algae, and bacteria. From the physical point of view, the first question is how to describe the absorption spectrum of LHC. To make a step towards the answer, here we consider one of the simplest light-harvesting (LH) complex, namely, LH2 complex of \textit{Rhodoblastus acidophilus}.

LH2 complexes of bacteria are popular model systems for investigation of light harvesting~\cite{Maity2023,Saga2022,Swainsbury2019,Frigaard1996,Anda2017,Qian2021,Segatta2017}. \textit{Rdb. acidophilus} LH2 complex consists of 18 altering $\alpha $ and $\beta $ protein chains, 9 rhodopin b-D-glucoside and 27 bacteriochlorophyll \textit{a} (BChl \textit{a}) molecules adopting C9 symmetry~\cite{Cherezov2006}. Bacteriochlorophyll molecules are arranged in two rings, one above another. Two characteristic peaks of LH2 forming the so-called $Q_y$ absorption band in the near infrared (NIR) region are associated with a closely (B850) packed ring of eighteen vertically aligned BChl \textit{a} and a loosely packed ring of nine BChl \textit{a} molecules perpendicular to the symmetry axis (B800). Here we focus on the ring of nine BChl \textit{a} molecules for a detailed investigation by the methods of the density functional theory (DFT). Comparison of the several DFT approaches with each other and with the experimental data allows us to choose (i) a minimal reliable model that includes BChl \textit{a} molecule and its surroundings and (ii) the most cost-effective DFT method for the description of the absorption spectra. This is a necessary step for further modeling of the whole LH2 complicated structure. We also consider the effect of Hartree-Fock exchange and demonstrate its importance for the proper description of bacteriochlorophyll-based structures spectral features.

\section{\label{Model} Computational methods and models}

For our modeling, we used 2FKW~\cite{RCSBPDB2FKW} PDB (Protein Data Bank) structure from the study of Cherezov et al.~\cite{Cherezov2006}.

Large protein-based structures like LH2 require the combination of state-of-the-art quantum chemical calculations with maybe less accurate but computationally inexpensive techniques. Previously, QM/MM (quantum mechanics/molecular mechanics) approach was implemented by Segatta et al.~\cite{Segatta2017} to simulate a linear and a two-dimensional electronic spectroscopy (2DES) spectra of \textit{Rhodoblastus acidophilus} LH2 complex. Some useful data were obtained from the semiempirical PM6 calculations~\cite{Tureli2011}. Here, we suggest hybrid DFT that is based on a GGA-pre-optimized structures for standalone BChl \textit{a} molecule and one-dimensional periodic models and DFTB method for further extended modeling.

Density functional-based tight binding (DFTB) method~\cite{Hourahine2020} gives reasonable results for a structure and optical spectra of chlorophylls and bacteriochlorophylls~\cite{Oviedo2010}. As this parametrized method is known to be effective in treating large-scale systems, it can be successfully used to model larger fragments of LH2, which are computationally too expensive when using DFT calculations. Third-order density functional-based tight binding (DFTB3) calculations were carried out in DFTB+ package using 3OB DFTB parameters~\cite{Gaus2013,Gaus2014,Lu2015} and Grimme D3 correction~\cite{Grimme2010,Grimme2011}.

DFT-GGA calculations were carried out in OpenMX (Open source package for Material eXplorer software package)~\cite{Boker2011} software using norm-conserving pseudopotentials~\cite{Bachelet1982,Troullier1991,Kleinman1982,Blochl1990,Morrison1993} and pseudoatomic orbitals (PAO)~\cite{Swainsbury2019,Frigaard1996,Anda2017,Qian2021} basis set. Further details of the calculations can be found in Supplementary Information (see Table~S1). Only the Perdew--Burke--Ernzerhof (PBE)~\cite{Perdew1996} exchange-correlation (XC) functional is available in the OpenMX package. Grimme D3 correction was used to account for the van-der-Waals interactions~\cite{Grimme2010,Grimme2011}.

We created the one-dimensional periodic structure with one BChl \textit{a} per unit cell to emulate the distance of 21.1~\AA~between BChl \textit{a} molecules in the B800 ring~\cite{Papiz2003}. Atomic structure and cell constant along the periodic direction were optimized within DFT-GGA approach. A vacuum interval of 20~\AA~was set and kept intact during the optimization in other directions to avoid artificial interactions between molecules in periodic boundary conditions (PBC). Monkhorst-Pack scheme~\cite{Monkhorst1976} was used for \textit{k}-point Brillouin zone sampling with the \textit{k}-mesh containing 1$\times$4$\times$1 points. $\Gamma(0,0,0)-Y(0,1/2,0)$ $k$-line was used for the band structure calculations procedure according to PBC formalism. Obviously, there is no actual band dispersion as we study molecule in vacuum and this calculation should be referred to molecular orbitals rather than bands in the solid state.

PBE XC functional is known to overestimate the electron delocalization effects and thus underestimate the band gap. Hybrid XC functionals like HSE06~\cite{Boker2011} and B3LYP~\cite{Stephens1994} include Hartree-Fock (HF) exchange improving the electronic structure near the Fermi level and the band gap values. $Q_y$ absorption peak, which is of the main interest here, lies in NIR range and corresponds to highest occupied molecular orbital-lowest unoccupied molecular orbital (HOMO-LUMO) transition. As this transition corresponds exactly to the band gap width, its correct description crucially depends on localization and energy of the frontier states, i.e. HOMO and LUMO.

HSE06 and B3LYP calculations of absorption spectra were carried out in VASP~\cite{Kresse1996} using OpenMX pre-optimized atomic structure due to the large size of the system and the computational cost of the hybrid DFT calculations. VASP uses plane wave (PW) basis set and projector augmented wave (PAW) method~\cite{Blochl1994} instead of the norm-conserving pseudopotentials. Figure~S1 demonstrates the comparison of spectra from GGA-PBE calculations in VASP and OpenMX. Both $Q_y$ peak positions and frontier orbitals localization (see Figure~S2) correspond well to each other.

Optical spectra were calculated within the linear response theory. Optical absorbance may be obtained from a complex dielectric function $\varepsilon(\omega)$ as:
\begin{equation}
 \varepsilon(\omega) = \varepsilon'(\omega) + \mathrm{i}\varepsilon''(\omega),
\end{equation}
where $\varepsilon'(\omega)$ and $\varepsilon''(\omega)$ are real and imaginary parts of the dielectric function, $\omega$ is the frequency. Real and imaginary parts of the dielectric function can be obtained as
\begin{equation}
 \varepsilon'(\omega) = 1 + \frac{2}{\pi} P\int\limits^{\infty}_0 {\frac{\varepsilon'(\omega')\omega'}{{\omega'}^2-\omega^2 + \mathrm{i}\eta}} d\omega',
\end{equation}
\begin{equation}
 \varepsilon''(\omega) = \frac{4\pi^2}{m^2\omega^2} \sum\limits_{V.C}{ \int\limits_{BZ}{d^3k \frac{2}{2\pi} {\left|e \cdot M_{CV}(k)\right|}^2 \times \Delta \left[E_C(k) - E_V(k) - \hbar\omega\right]}},
\end{equation}
where $k$ is the reciprocal cell vector, $\hbar\omega$ is the photon energy, $M_{CV}(k)$ is an element of a dynamical matrix, BZ means integrating within the Brillouin zone, $C$ and $V$ stand for conduction and valence band.

Optical absorption coefficient $\alpha$ may be calculated as
\begin{equation}
 \alpha(\omega) = \sqrt{2\omega} + \sqrt{\sqrt{\varepsilon'^2(\omega)-\varepsilon''^2(\omega)}-\varepsilon'(\omega)}.
\end{equation}

\section{\label{Results} Results and discussion}

\begin{figure}
\centering
 \includegraphics[width=0.75\linewidth]{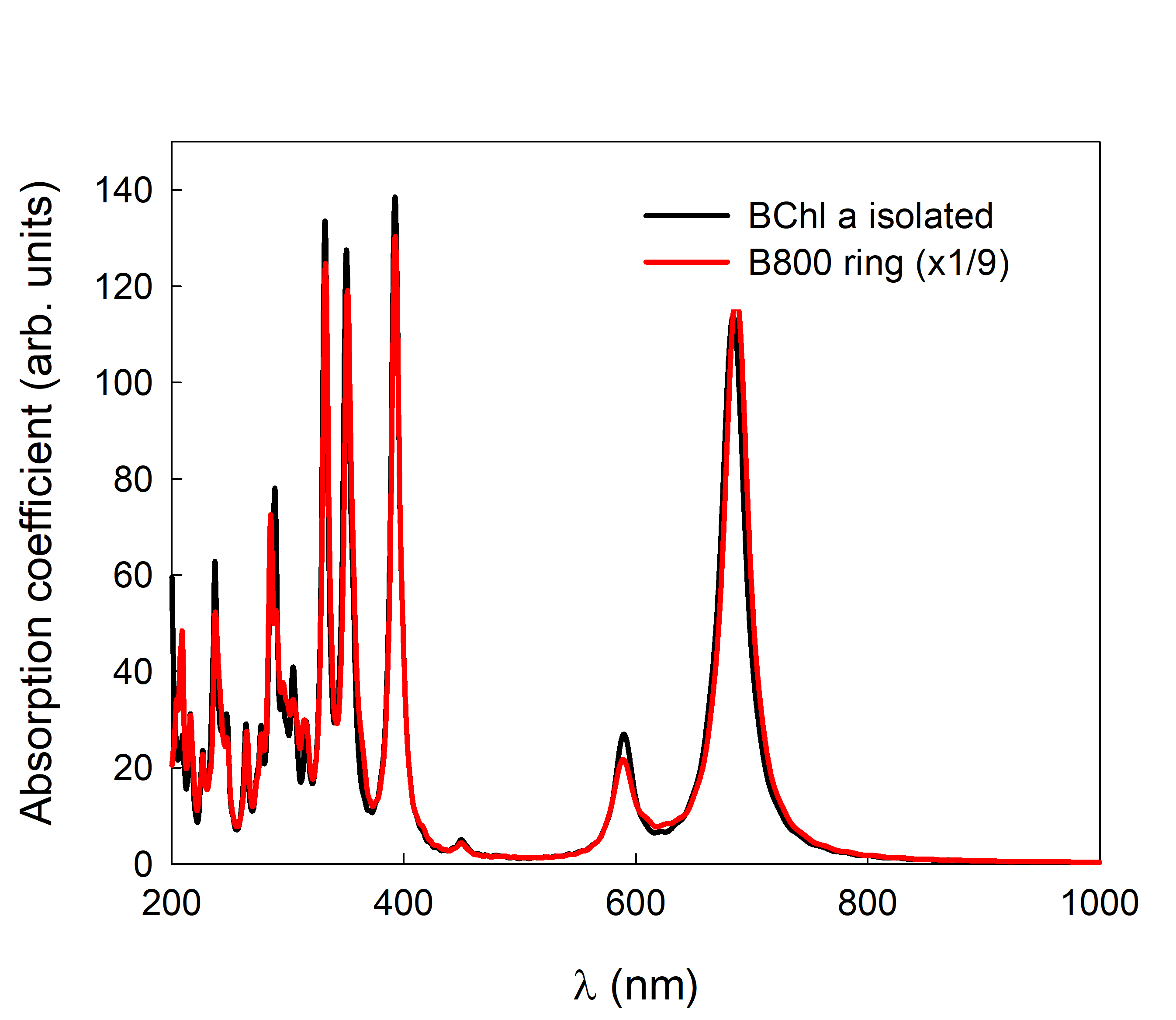}
 \caption{\label{Fig1} Absorption spectra of the isolated BChl \textit{a} molecule (black) and the B800 ring containing nine BChl \textit{a} molecules extracted from the LH2 complex (red).}
\end{figure}

As the first step, we compare the absorption spectrum of the whole B800 ring with that of the isolated BChl \textit{a} molecule. Here we use DFTB method for geometry optimization and absorption spectra calculations since its parameterization for the bacteriochlorophylls is shown to give results~\cite{Oviedo2010} in good agreement with the experimental spectrum. Both BChl \textit{a} and B800 ring were fully optimized till the forces acting on atoms were less than $10^{-4}$~Hartree/Bohr. The calculated absorption spectra for the ring containing all nine BChl \textit{a} molecules together with the isolated BChl \textit{a} are shown in Figure~\ref{Fig1}. There are two prominent NIR peaks, $Q_x$ (590 nm) and $Q_y$ (683 nm) and the HOMO-LUMO gap was estimated to be 1.14 eV. Absorption peaks are the same for both the ring and the isolated molecule, thus the interaction between the neighboring molecules in the ring is vanishingly small. Therefore, we can safely focus on the isolated BChl \textit{a} molecule to study the role of the interactions and the surroundings on the spectrum.

\subsection{\label{sec3.1} BChl \textit{a} molecule electronic structure and optical spectrum from GGA-PBE calculations}

\begin{figure}
\centering
 \includegraphics[width=0.95\linewidth]{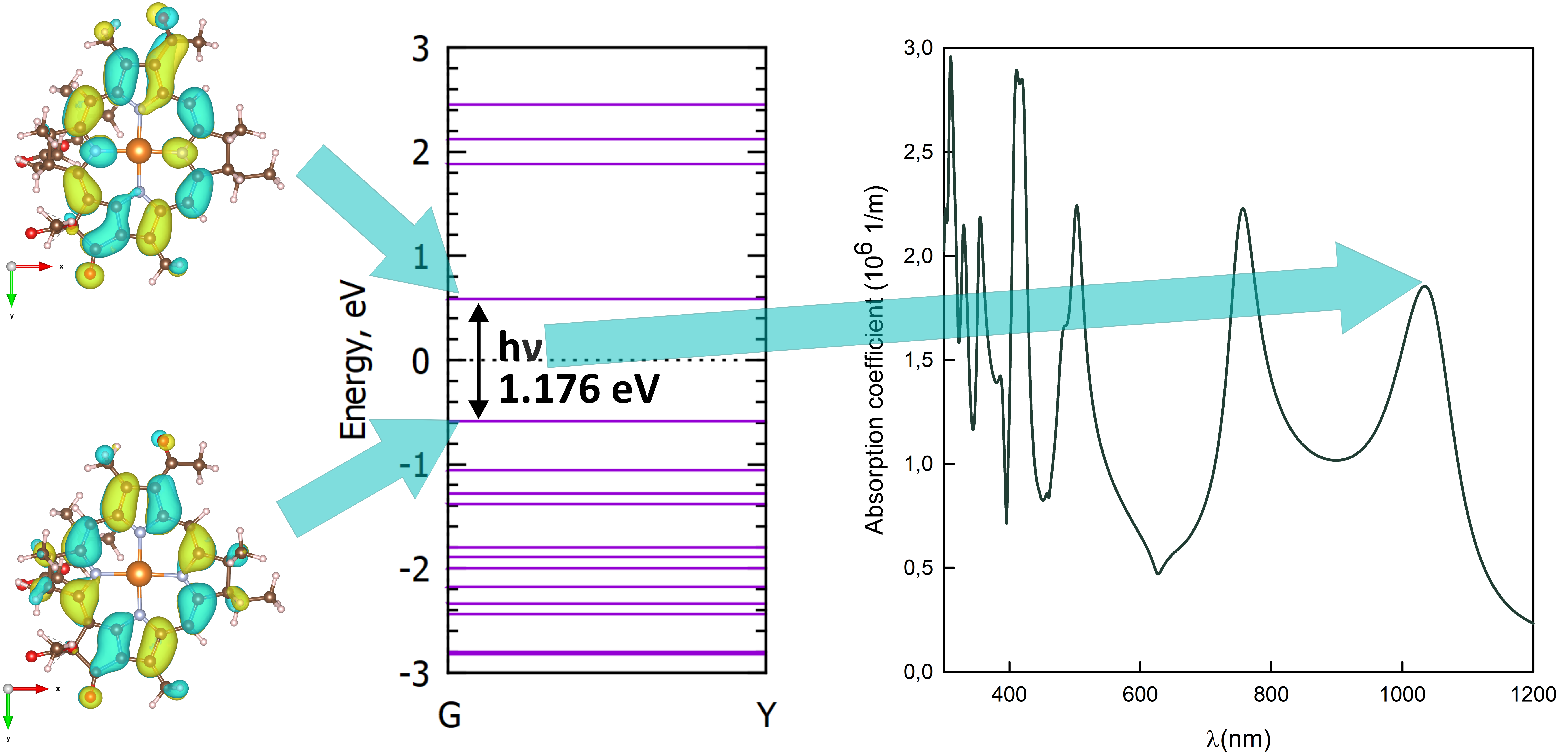}
 \caption{\label{Fig2} HOMO and LUMO localization, energy gap and corresponding absorption peak from GGA-PBE/PAO calculation in OpenMX package. Mg, C, N, O and H are represented by orange, brown, blue, red and pink balls, respectively. Phytyl tail is truncated. Yellow and blue areas correspond to the sign of the wavefunction, isosurface level is set to 0.02 $\alpha_0^{-3}$, $\alpha_0$ is the Bohr radius.}
\end{figure}

As our PBC model is constructed to emulate experimentally observed distance between BChl \textit{a} in B800 ring, unit cell optimization was performed giving a distance of 20.536~\AA~between the neighboring BChl \textit{a} molecules that is in a good agreement with the XRD data~\cite{RCSBPDB2FKW}. Though there is a formal structural periodicity, in the absence of any surroundings, here we can consider BChl \textit{a} as an isolated molecule. However, the molecule itself adopts perfect planar structure after the optimization while in B800 its porphyrin ring is visibly distorted forming a dome-like structure due to the interaction with surrounding amino acids. Figure~\ref{Fig2} demonstrates the absorption spectrum calculated for the optimized structure of BChl \textit{a}. NIR absorption peak associated with the $Q_y$ band is located at 1034 nm. It corresponds to the HOMO-LUMO transition with the energy gap of 1.176 eV. Left panel of Figure~\ref{Fig2} presents the localization of the frontier orbitals. $Q_x$ peak is located at 751 nm (see Table~\ref{table1}).

\begin{table}[H]
\caption{$Q_x$ and $Q_y$ absorption peak positions from GGA-PBE/PAO calculations with respect to amino acids in structure, nm. The minimal reliable model is highlighted.}
\label{table1}
\begin{center}
\smallskip {\setlength{\tabcolsep}{2pt}
\begin{tabular}{@{\extracolsep{0.15in}}p{0.7in}p{0.4in}p{0.6in}p{0.6in}>{\columncolor[gray]{0.9}}p{0.6in}p{0.6in}p{0.6in}}
\toprule
Structure & BChl \textit{a} & BChl \textit{a} + Cxm-Asn-Gln & BChl \textit{a} + Cxm +Arg & BChl \textit{a} + Cxm-Asn +Arg & BChl \textit{a} + Cxm-Asn-Gln +Arg & BChl \textit{a} + Cxm-Asn-Gln +Arg-Thr \\
\hline\noalign{\smallskip}
Location of $Q_x$ maximum, nm & 751 & 776 & 792 & 798 & 794 & 798 \\
\hline
Location of $Q_y$ maximum, nm & 1034 & 1044 & 1053 & 1060 & 1058 & 1060 \\ \bottomrule
\end{tabular}
}
\end{center}
\end{table}

\subsection{\label{sec3.2} Amino acids influence on BChl \textit{a} absorption spectrum from GGA-PBE calculations}

Pigment interactions with protein chains are known to affect its structure~\cite{Saga2022,Anda2017,Qian2021}. Thus, we investigated the amino acid surroundings impact on BChl \textit{a} absorption spectrum creating one-dimensional model containing two blocks of  acids which are in direct contact with BChl \textit{a} molecule. First one has three $\alpha $ chain amino acids located above the B800 plane in LH2, namely, N-carboxymethionine (Cxm), asparagine (Asn), and glutamine (Gln). Second block contains two side-located $\beta $ chain amino acids: arginine (Arg) and threonine (Thr).

\begin{figure}
\centering
 \includegraphics[width=0.75\linewidth]{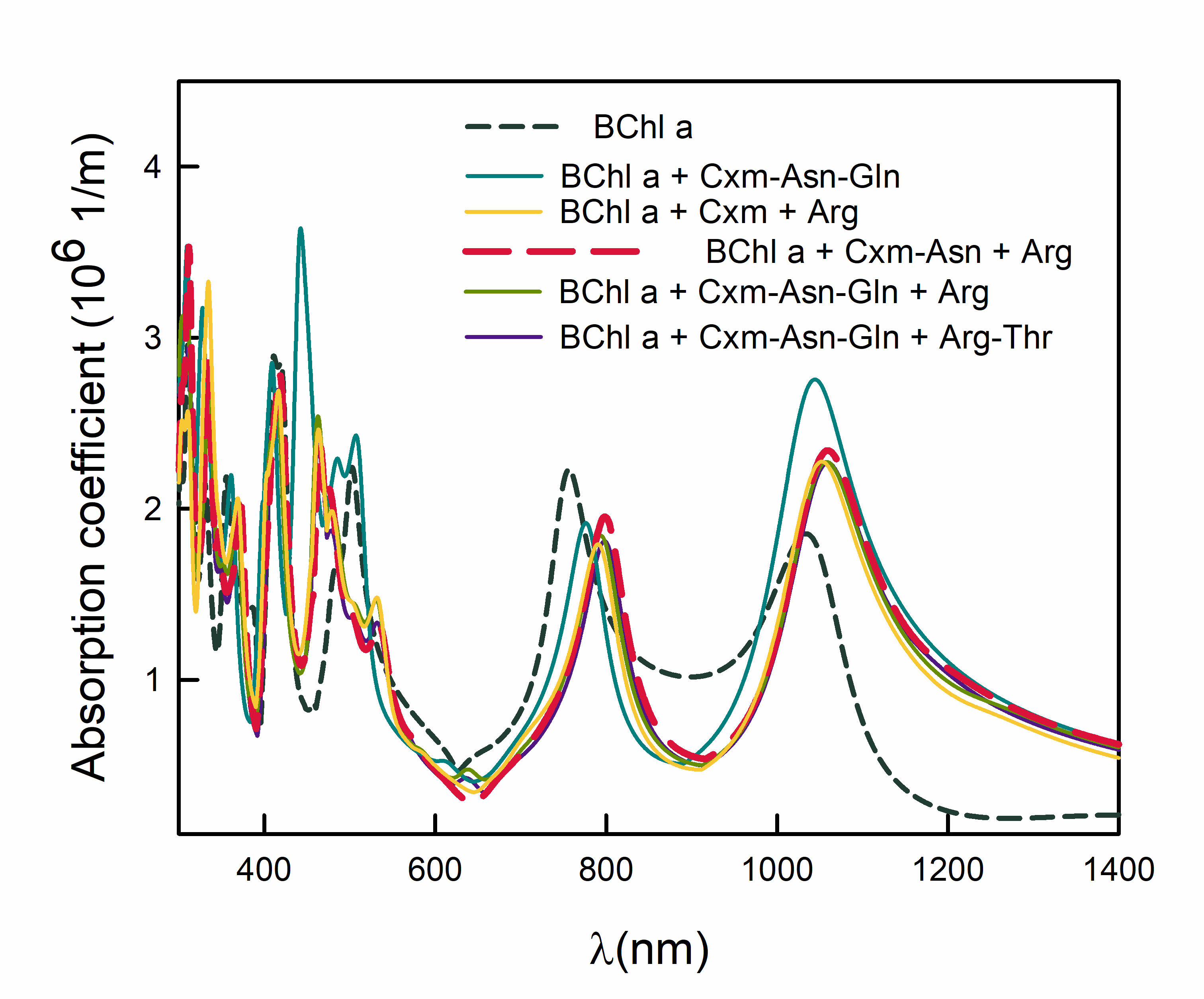}
 \caption{\label{Fig3} Absorption spectra of the isolated BChl \textit{a} and aminoacids containing structures from GGA-PBE/PAO calculations in OpenMX. N-carboxymethionine, asparagine, glutamine, arginine and threonine denoted as Cxm, Asn, Gln, Arg and Thr.}
\end{figure}

Since we are interested in finding a minimal model, i.e. a model that includes the BChl \textit{a} molecule and a minimal yet necessary subset of the surrounding molecules, we include the mentioned amino acids one by one and compare the results. Figure~\ref{Fig3} illustrates absorption spectra for studied subsets of the BChl \textit{a} surrounding. Both $Q_x$ (751 nm) and $Q_y$ (1034 nm) peaks are red shifted in all cases. N-carboxymethionine, asparagine, and arginine are found to be mostly responsible for this, see Table~\ref{table1}. It is in agreement with the data regarding Mg chelation by carboxylated N-terminus of the $\alpha$ chain and the enhancement of electron delocalization due to the acetyl group rotation when interacting with arginine~\cite{Swainsbury2019,Tureli2011}, which leads to the to the longer wavelength (bathochromic shift) and absorption band narrowing. Besides, an additional absorption peak at 563 nm appears for all arginine-containing structures.

Further expansion of the model by adding glutamine above the porphyrin ring and threonine alongside the molecule does not affect much neither $Q_y$ nor $Q_x$ peak position, see Table~\ref{table1}. Therefore, the abovementioned structure including BChl \textit{a} molecule and three amino acids is determined as the minimal model sufficient for the description of B800 ring of LH2 complex and will be used in further hybrid and DFTB calculations.

\subsection{\label{sec3.3} Methods comparison}

\begin{figure}
\centering
 \includegraphics[width=0.75\linewidth]{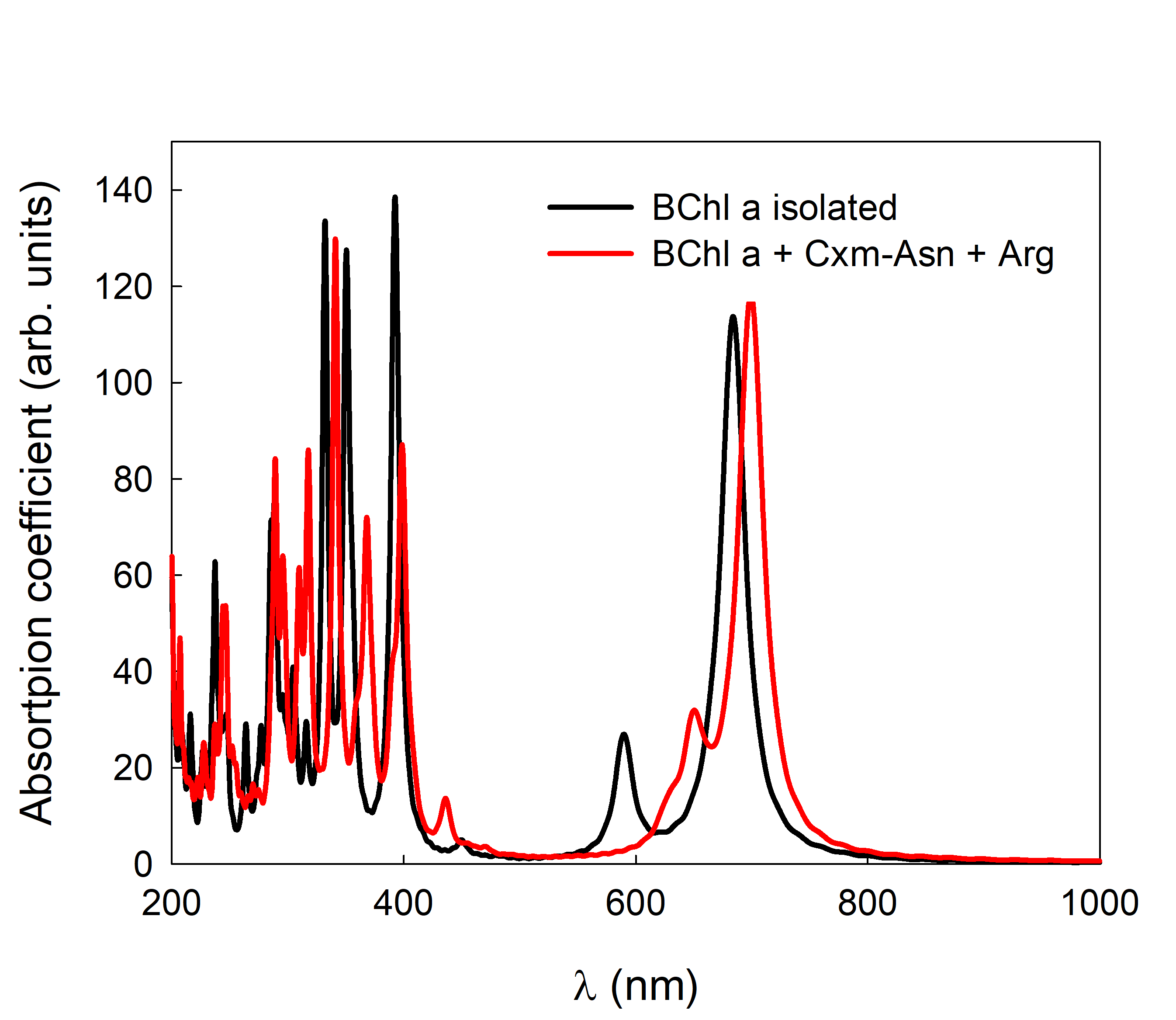}
 \caption{\label{Fig4} Absorption spectra of isolated BChl \textit{a} molecule and BChl~+~Cxm-Asn~+~Arg structure from DFTB calculations.}
\end{figure}

\begin{figure}
\centering
 \includegraphics[width=0.85\linewidth]{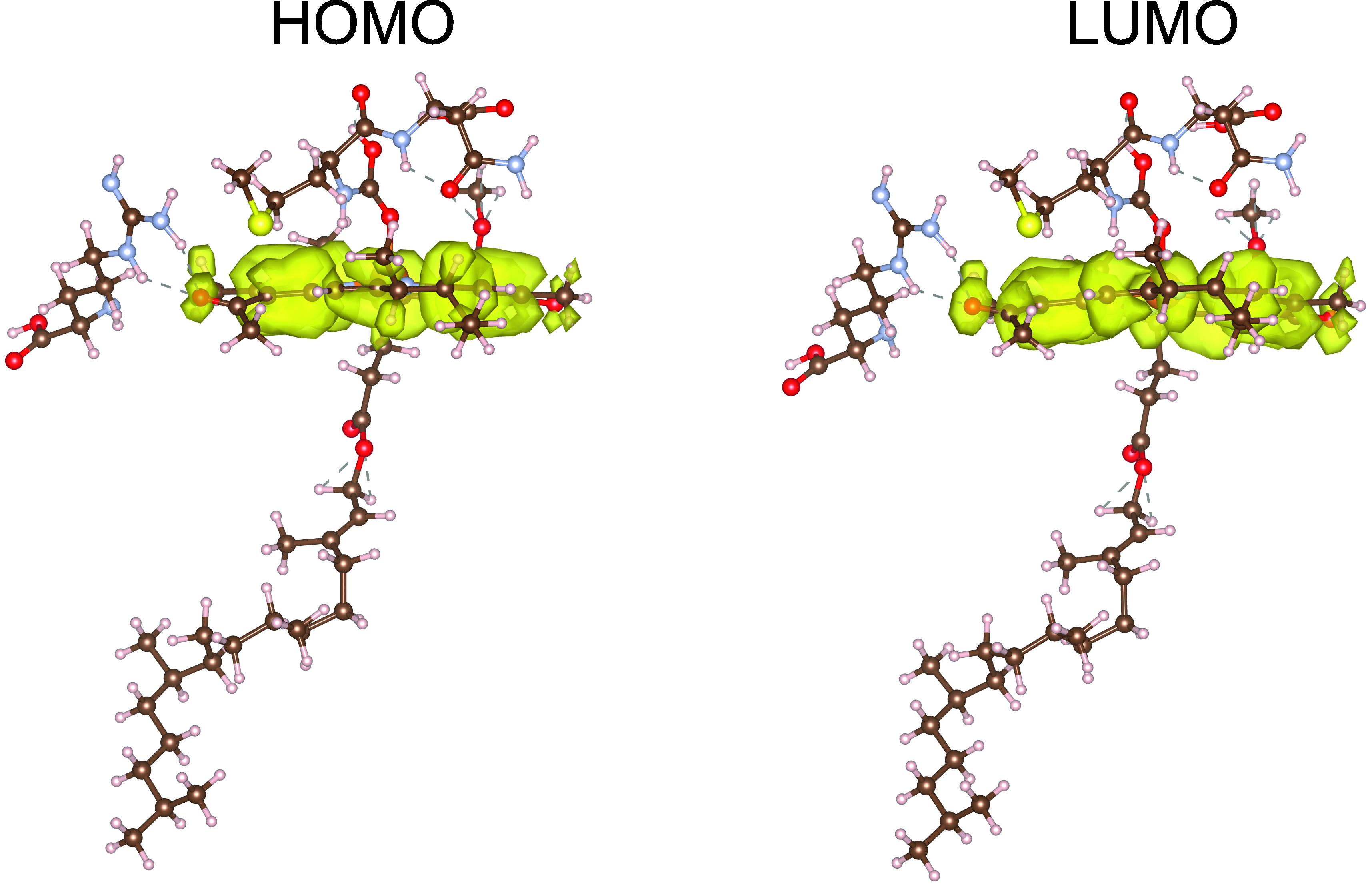}
 \caption{\label{Fig5} HOMO and LUMO spatial distribution of electron density of BChl~+~Cxm-Asn~+~Arg structure from DFTB calculation. Mg, C, N, O and H are represented as orange, brown, blue, red and pink balls, respectively.}
\end{figure}

DFTB method shows the same trend for the bathochromic shift of both $Q_x$ and $Q_y$ absorption peaks (see Figure~\ref{Fig4}) and additional arginine-associated peak appearance when the model is augmented with three amino acids defined as the minimal reliable model. According to the spatial electron density distribution of HOMO and LUMO orbitals shown in Figure~\ref{Fig5}, the presence of amino acids does not affect the shape of the frontier orbitals, though shifting their energy. Indeed, electron density is virtually the same as one of isolated BChl \textit{a} molecule and localizes mostly on porphyrin ring carbon (HOMO) and nitrogen (LUMO) atoms.

\begin{figure}
\centering
 \includegraphics[width=0.75\linewidth]{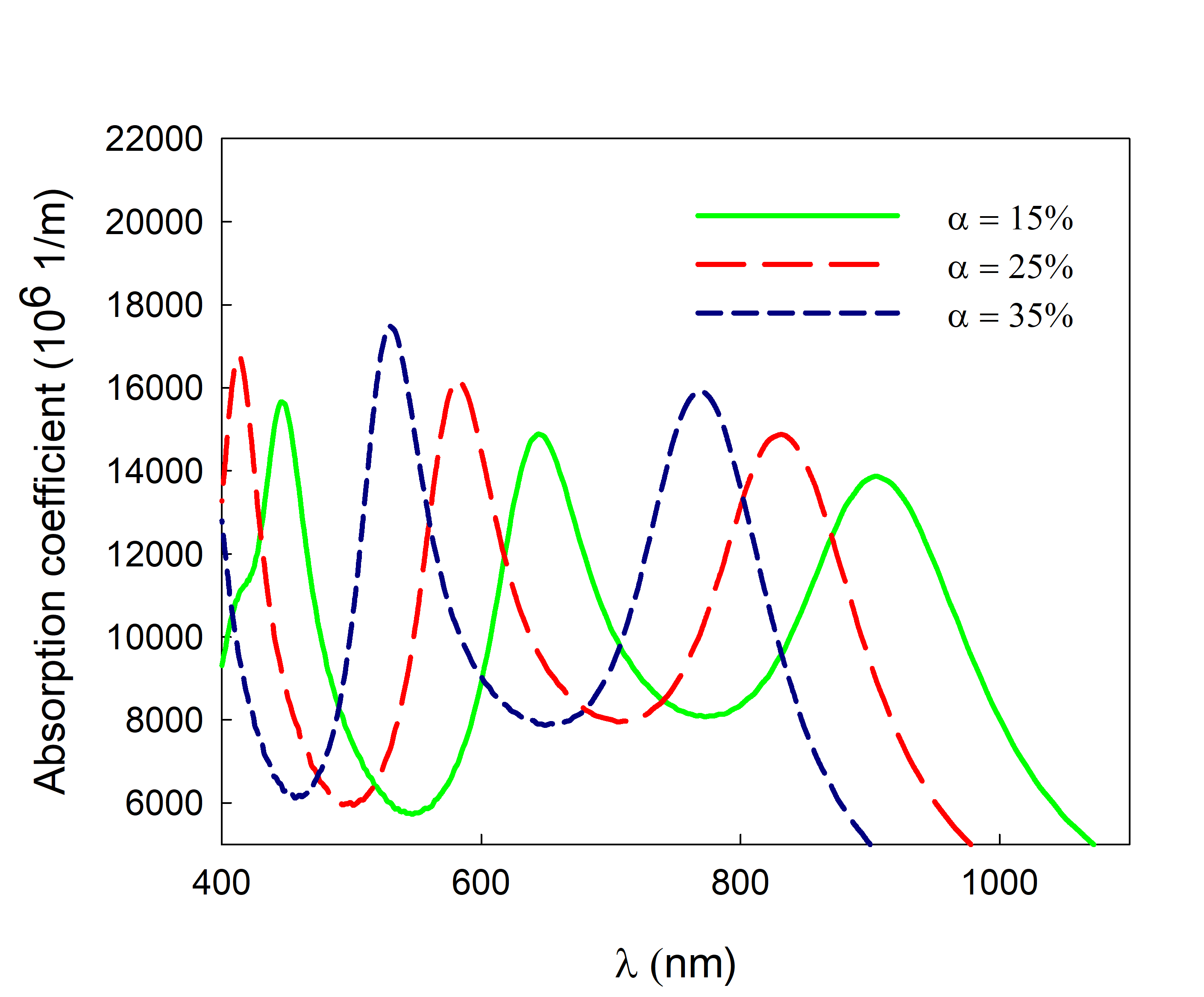}
 \caption{\label{Fig6} Absorption spectra of the isolated BChl \textit{a} molecule from GGA-PBE and hybrid B3LYP or HSE06 calculations in VASP using the PW basis.}
\end{figure}

\begin{figure}
\centering
 \includegraphics[width=0.95\linewidth]{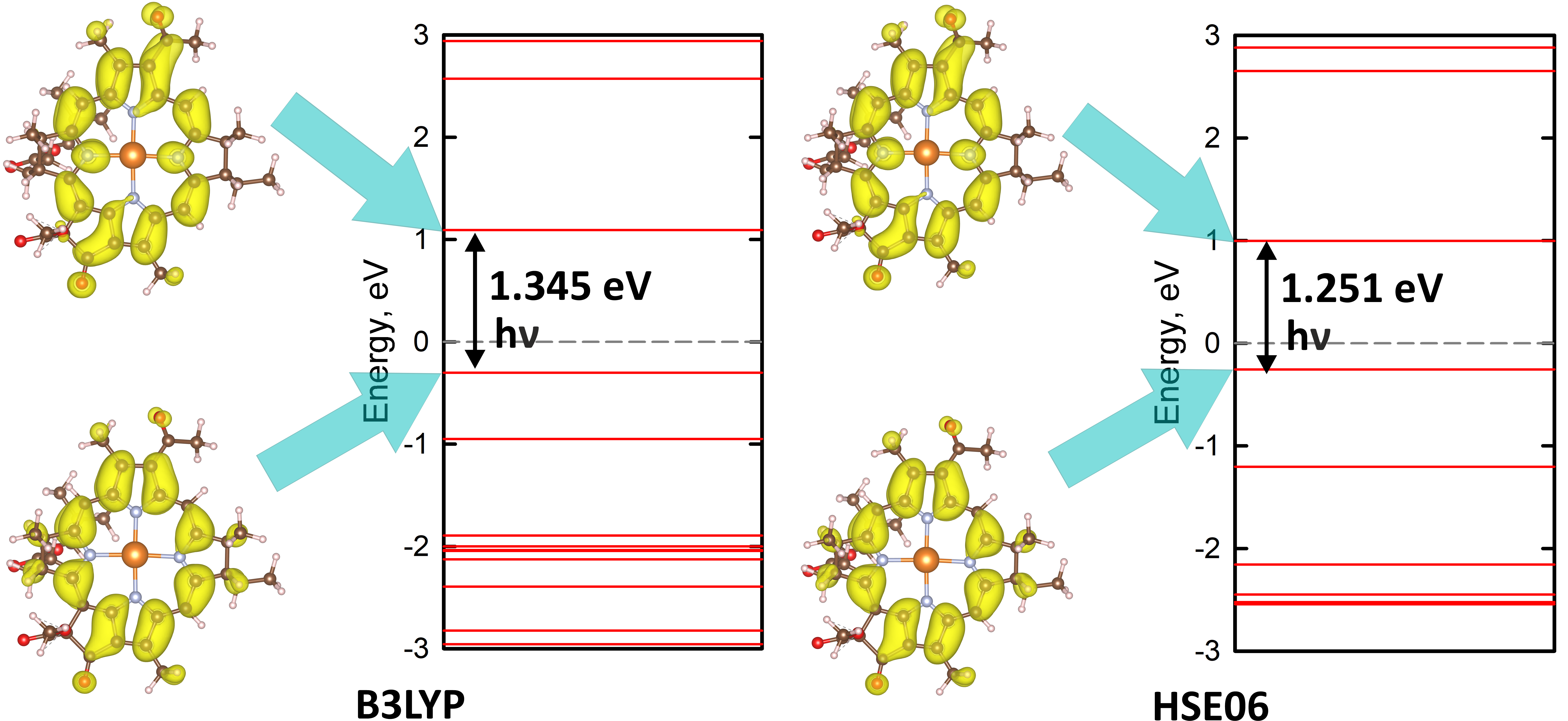}
 \caption{\label{Fig7} HOMO and LUMO spatial distribution of electron density, energy gap from hybrid XC functionals PW calculation in VASP package. Mg, C, N, O and H are represented as orange, brown, blue, red and pink balls, respectively. Phytyl tail is truncated. Isosurface level is set to 0.00085$\alpha_0^{-3}$, $\alpha_0$ is the Bohr radius.}
\end{figure}

\begin{figure}
\centering
 \includegraphics[width=0.75\linewidth]{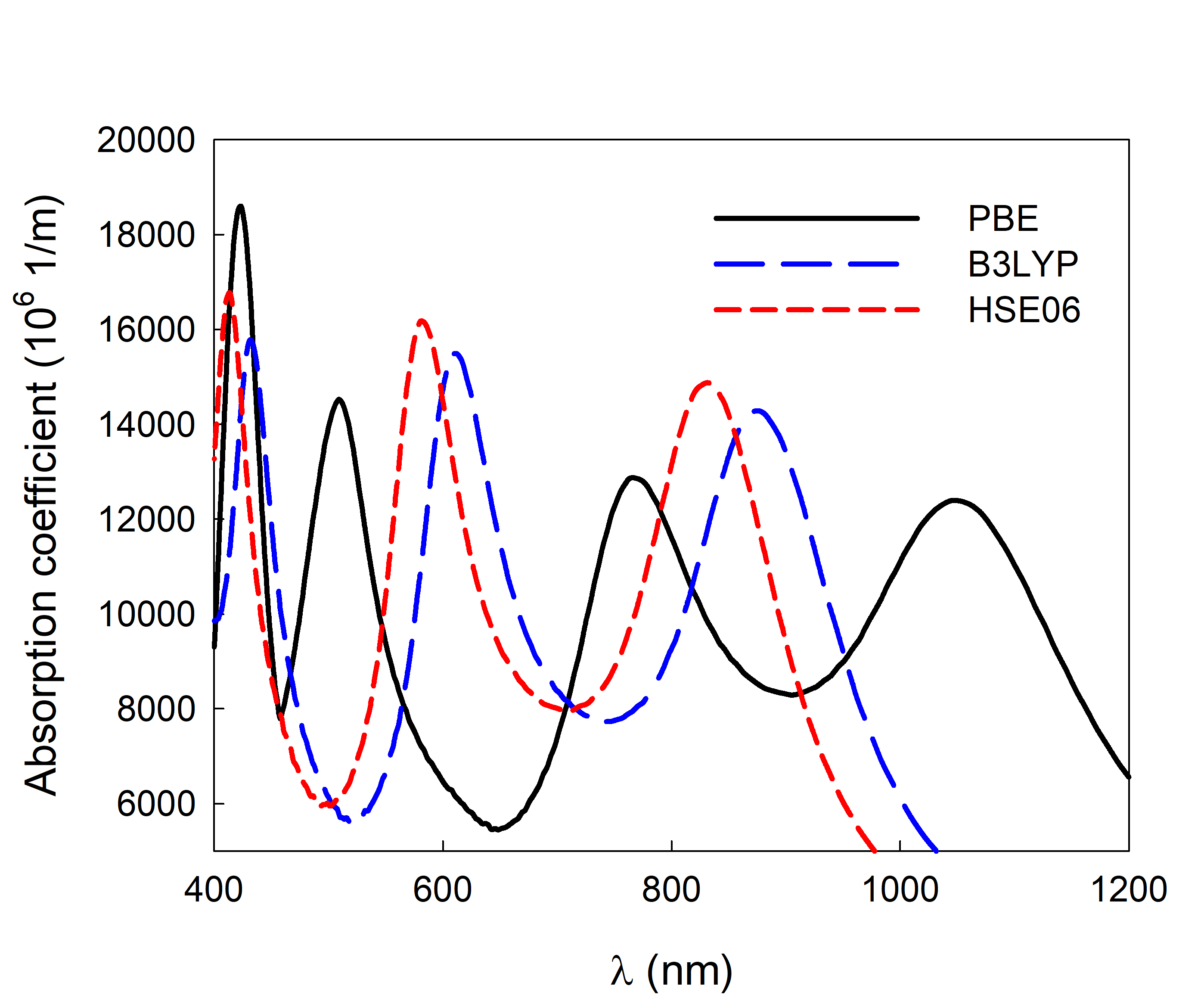}
 \caption{\label{Fig8} Absorption spectra of the isolated BChl \textit{a} molecule from HSE06 calculations with respect to the share of HF exchange proportional to the parameter~$\alpha$.}
\end{figure}

Inclusion of the HF exchange in DFT calculations widens the HOMO-LUMO gap and blue-shifts the absorption peak with respect to GGA-PBE, see Figure~\ref{Fig6}. B3LYP and HSE06 result in the frequency of 876 nm and 831 nm for the $Q_y$ peak, respectively, $Q_x$ band is blue-shifted as well (see Table~\ref{table2}). Frontier orbitals localization, however, stays the same, see Figure~\ref{Fig7}. HOMO-LUMO gaps are 1.345 eV and 1.251 eV for B3LYP and HSE06, respectively. To study the role of the HF exchange, we vary the amount of its share denoted by $\alpha$ in HSE06 calculations by setting $\alpha=15$, $25$ and $35$\%. As seen in Figure~\ref{Fig8} (see exact peak values in Table~S2), the shift of the peak to the smaller wavelength (hypsochromic shift) increases with the increase of $\alpha$. For a large amount of HF share, $\alpha=35$\%, both $Q_x$ and $Q_y$ peaks positions are in good agreement with ones obtained by DFTB. Since the DFTB parameterization was optimized to agree with B3LYP calculation results and thus contains some part of HF exchange implicitly as well, we conclude that it's extremely important for the proper description of the BChl \textit{a} molecule spectra.

\begin{table}[H]
\caption{$Q_x$ and $Q_y$ absorption peak position for the isolated BChl \textit{a} molecule and in the minimal one-dimensional model of the B800 ring calculated via different DFT functionals and DFTB method (shift with respect to PBE is shown in brackets), nm.}
\label{table2}
\begin{center}
\smallskip {\setlength{\tabcolsep}{2pt}
\begin{tabular}{p{1.6in}p{0.8in}p{0.8in}p{0.8in}p{0.8in}}
\toprule
Structure & PBE & B3LYP & HSE06 & DFTB \\
\hline\noalign{\smallskip}
BChl & 767, 1046 & 612, 876\newline (-155, -170) & 581, 831\newline (-186, -215) & 590, 683\newline (-177, -363) \\
\hline
BChl + Cxm-Asn +Arg & 790, 1072 & 627, 885\newline (-163, -187) & 609, 855\newline (-181, -217) & 651, 700\newline (-139, -372) \\
\bottomrule
\end{tabular}
}
\end{center}
\end{table}

Table~\ref{table2} illustrates virtually the same hypsochromic shift (\~{}216 nm) of $Q_y$ peak observed for HSE06 XC functional with respect to GGA-PBE for both isolated molecule and our amino acid model. This may be used as a guide for brief estimation of absorption maximum position in further studies by adding corresponding value to the ones obtained from the less time consuming PBE calculations.

\section{\label{Conclusion} Conclusion}

We modeled B800 part of \textit{Rdb. acidophilus} LH2 complex using various DFT-based techniques. Absorption peaks are the same for both the B800 ring and the isolated BChl \textit{a} molecule according to DFTB calculations, thus the interaction between the neighboring bacteriochlorophyll molecules in the ring is negligible. We have determined three amino acids in $\alpha $ (Cxm, Asn) and $\beta $ (Arg) protein chains that have maximal effect on the B800 absorption peak in \textit{Rdb. acidophilus} LH2 complex. Thus, accompanying the BChl \textit{a} molecule, they form a minimal reliable model for further description of the LH2. Including Hartree-Fock exchange provides hypsochromic shift, the more prominent the more its share is, as we've shown in our HSE calculations. Thus, we show the importance of this Coulomb interactions part for the formation of the absorption spectra features. Hybrid HSE06 DFT functional demonstrates the same shift of characteristic B800 peak with respect to GGA for both isolated BChl \textit{a} molecule and pigment-protein complex. Therefore, using this shift one can estimate the absorption peak position from PBE calculations in extended models as well. DFTB method proved to give reasonable results for B800 part of LH2 complex and thus may be used to simulate other large fragments within the light-harvesting structure.

\supplementary{
The following are available as Supplementary Materials,
Figure~S1: Absorption spectra of BChl \textit{a} molecule from GGA-PBE calculations with either PAO (OpenMX) or PW (VASP) basis set used. Figure~S2: HOMO and LUMO spatial distribution of electron density, energy gap and corresponding absorption peak from GGA-PBE/PW calculation in VASP package. Mg, C, N, O and H are represented as orange, brown, blue, red and pink balls, respectively. Isosurface level is set to 0.00085$\alpha_0^{-3}$, $\alpha_0$ is Bohr radius. Table~S1: Computational parameters defined as sufficient for correct description of BChl \textit{a}. Table~S2: $Q_y$ absorption peak position for BChl \textit{a} molecule with respect to $\alpha $ share of Hartree-Fock exchange in HSE functional.}


\authorcontributions{Conceptualization and supervision, V.F.S. and M.M.K.; data curation and visualization, L.V.B. and E.A.K.; formal analysis, E.A.K.; writing, L.V.B., E.A.K. and M.M.K. All authors have read and agreed to the published version of the manuscript.}

\funding{This work was supported by the state assignment of the Ministry of Science and Higher Education of the Russian Federation.}

\acknowledgments{Authors would like to thank Information Technology Centre, Novosibirsk State University for providing access to their supercomputers. L.V.B.  would like to thank Irkutsk Supercomputer Center of SB RAS for providing the access to HPC-cluster «Akademik V.M. Matrosov» (Irkutsk Supercomputer Center of SB RAS, Irkutsk: ISDCT SB RAS; http://hpc.icc.ru, accessed 20.10.2023)}

\conflictsofinterest{The authors declare no conflict of interest.}

\begin{adjustwidth}{-\extralength}{0cm}

\reftitle{References}






\end{adjustwidth}

\clearpage

\setcounter{equation}{0}
\setcounter{figure}{0}
\setcounter{table}{0}
\setcounter{page}{1}
\setcounter{section}{0}
\makeatletter
\renewcommand{\theequation}{S\arabic{equation}}
\renewcommand{\thefigure}{S\arabic{figure}}
\renewcommand{\thetable}{S\arabic{table}}
\renewcommand{\thesection}{S-\Roman{section}}
\renewcommand{\bibnumfmt}[1]{[S#1]}
\renewcommand{\citenumfont}[1]{S#1}

\section{\large Supplementary Materials: ``Absorption spectra of the purple nonsulfur bacteria light-harvesting complex: a DFT study of the B800 part'' \label{Sec.Suppl}}

Here we provide some technical notes on the calculations in OpenMX package. The cutoff energy value was equal to 150~Ry. All structures were relaxed until forces acting on the atoms became less than $10^{-5}$~Hartree per Bohr. The convergence condition for the energy minimization was equal to $10^{-7}$~Hartree.

\begin{table}[H]
\caption{Computational parameters defined as sufficient for correct description of BChl \textit{a}.}
\label{tableS1}
\begin{center}
\smallskip {\setlength{\tabcolsep}{2pt}
\begin{tabular}{p{1.4in}p{1.5in}p{1.5in}}
\toprule
Element & PAO basis set & Cutoff radius \\
\hline\noalign{\smallskip}
Mg & s2p3d2 & 9.0 \\
C & s2p2d2 & 5.0 \\
N & s2p2d2 & 5.0 \\
O & s2p2d2f1 & 5.0 \\
H & s2 & 6.0 \\
\bottomrule
\end{tabular}
}
\end{center}
\end{table}

The convergence condition for the energy minimization in VASP was equal to $10^{-7}$~eV, plane wave cutoff energy was set to 400~eV.

\begin{figure}[H]
\centering
 \includegraphics[width=0.75\linewidth]{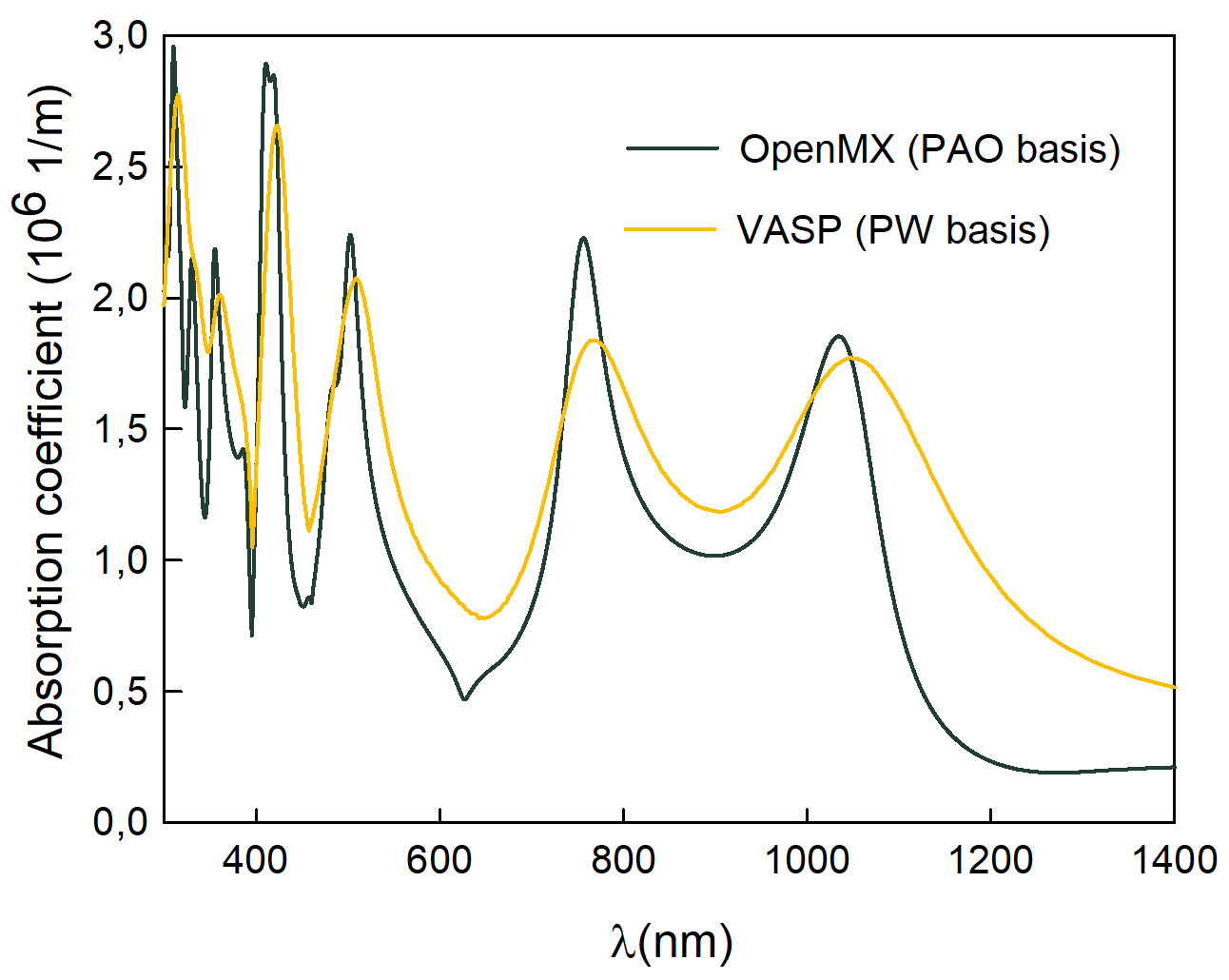}
 \caption{\label{FigS1} Absorption spectra of BChl \textit{a} molecule from GGA-PBE calculations with either PAO (OpenMX) or PW (VASP) basis set used.}
\end{figure}

\begin{figure}[H]
\centering
 \includegraphics[width=0.95\linewidth]{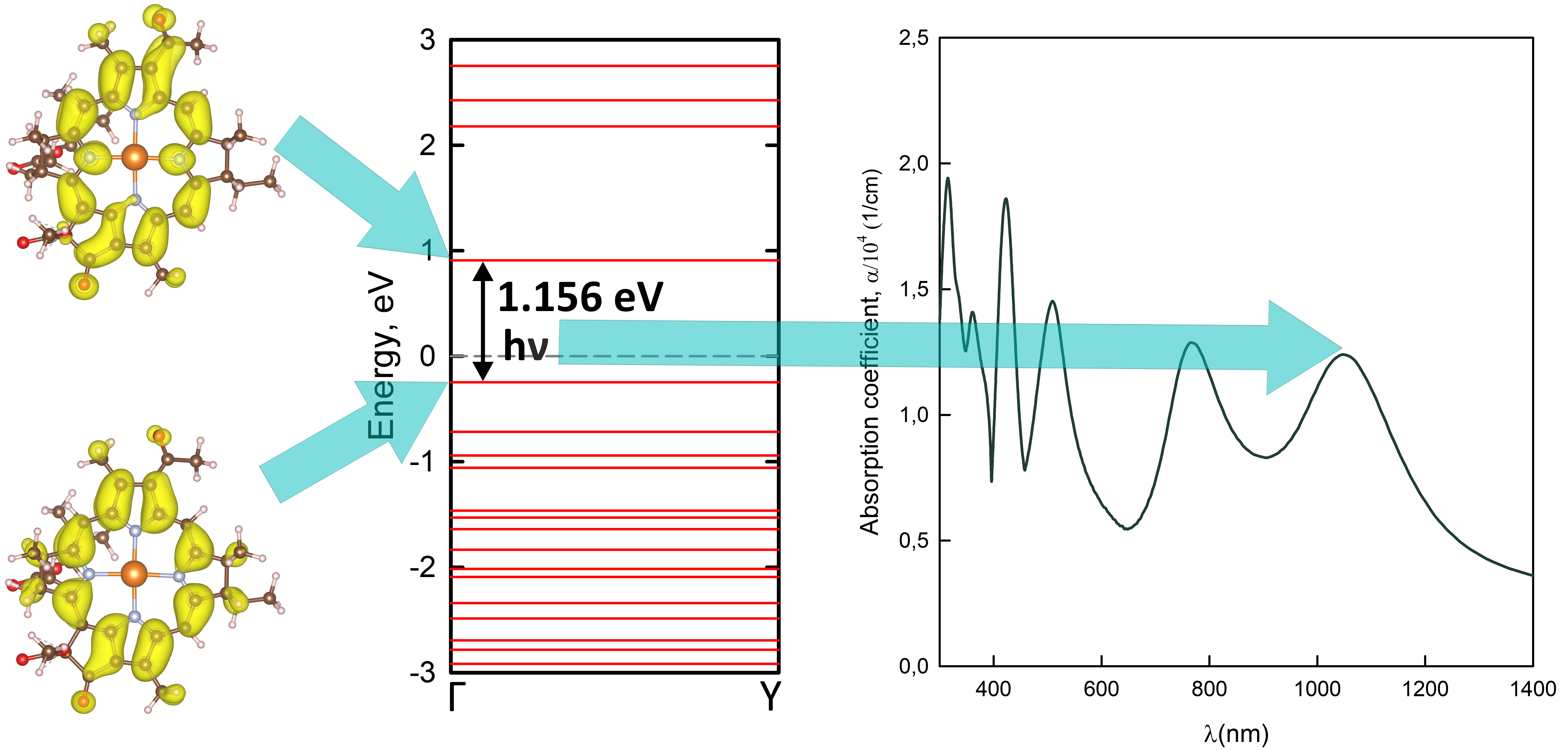}
 \caption{\label{FigS2} HOMO and LUMO spatial distribution of electron density, energy gap and corresponding absorption peak from GGA-PBE/PW calculation in VASP package. Mg, C, N, O and H are represented as orange, brown, blue, red and pink balls, respectively. Isosurface level is set to 0.00085$\alpha_0^{-3}$, $\alpha_0$ is Bohr radius.}
\end{figure}

\begin{table}[H]
\caption{$Q_y$ absorption peak position for BChl \textit{a} molecule with respect to $\alpha $ share of Hartree-Fock exchange in HSE functional.}
\label{tableS2}
\begin{center}
\smallskip {\setlength{\tabcolsep}{2pt}
\begin{tabular}{p{1.1in}p{0.7in}p{1.8in}p{0.7in}}
\toprule
$\alpha $, \% & 15 & 25 (default HSE06 value) & 35 \\
\hline\noalign{\smallskip}
$Q_y$, nm & 905 & 831 & 769 \\
\bottomrule
\end{tabular}
}
\end{center}
\end{table}

\end{document}